\documentclass[a4paper,notitlepage,nobibnotes,nofootinbib,superscriptaddress]{revtex4-1}

\usepackage{amssymb}
\usepackage{amsmath}
\usepackage{epsfig}

\newcommand{\be}{\begin{equation}}
\newcommand{\ee}{\end{equation}}
\newcommand{\bea}{\begin{eqnarray}}
\newcommand{\eea}{\end{eqnarray}}

\newcommand{\g}{\gamma}
\newcommand{\f}{\frac}

\newcommand{\intc}[1]{{\int\frac{d#1}{2i\pi}}}
\newcommand\lr[1]{{\left({#1}\right)}}

\begin{document}


\title{Gaps between jets in double-Pomeron-exchange processes at the LHC}

\author{C. Marquet}\email{cyrille.marquet@cern.ch}
\affiliation{Departamento de F\'\i sica de Part\'\i culas and IGFAE,  Universidade de Santiago de Compostela, 15782 Santiago de Compostela, Spain}
\affiliation{Physics Department, Theory Unit, CERN, 1211 Gen\`eve 23, Switzerland}
\author{C. Royon}\email{christophe.royon@cea.fr}
\affiliation{IRFU/Service de physique des particules, CEA/Saclay, 91191 Gif-sur-Yvette cedex, France}
\author{M. Trzebi\'nski}\email{maciej.trzebinski@cern.ch}
\affiliation{Institute of Nuclear Physics PAN, ul. Radzikowskiego 152, 31-342 Krak\'ow, Poland}
\author{R. \v{Z}leb\v{c}\'{i}k}\email{zlebcr@mail.desy.de}
\affiliation{Faculty of Mathematics and Physics, Charles University, Praha, Czech Republic}

\begin{abstract}

The possibility to measure jet-gap-jet final states in double-Pomeron-exchange events at the LHC is presented.
In the context of the ATLAS experiment with additional forward physics detectors, cross sections for different experimental
settings and gap definitions are estimated. This is done in the framework of the Forward Physics Monte Carlo interfaced
with a perturbative QCD model that successfully reproduces standard jet-gap-jet cross sections at the Tevatron. The
extrapolation to LHC energies follows from the Balitsky-Fadin-Kuraev-Lipatov dynamics, implemented in the model at
next-to-leading logarithmic accuracy.

\end{abstract}

\maketitle

\section{Introduction}

In a hadron-hadron collision, a jet-gap-jet event features a large rapidity gap with a high-$p_T$ jet on each side ($p_T\!\gg\!\Lambda_{QCD}$). Across the gap, the object exchanged in the $t-$channel is color singlet and carries a large momentum transfer. When the rapidity gap is sufficiently large, which requires a large collision energy ($\sqrt{s}\gg p_T$), the perturbative QCD description of jet-gap-jet events boils down to the exchange of a Balitsky-Fadin-Kuraev-Lipatov (BFKL) Pomeron \cite{bfkl,nllbfkl}. The first measurements carried out at the Tevatron \cite{d0,cdf} allowed preliminary tests of that QCD description, which were quite successful \cite{Chevallier:2009cu,Kepka:2010hu}, paving the way for more thorough investigations at the LHC.

In \cite{Chevallier:2009cu}, the full next-to-leading logarithm (NLL) BFKL kernel was implemented including all conformal spins and necessary collinear improvements. In \cite{Kepka:2010hu}, the resulting parton-level hard cross section was interfaced with the HERWIG Monte Carlo \cite{herwig}, as was first proposed in \cite{cfl} in the case of the leading-logarithm (LL) BFKL calculation, in order to take into account parton showering, hadronization, and jet reconstruction effects. The resulting model was able to reproduce the CDF and D0 data, and to make predictions for the LHC, where many such events are expected to be produced.

Unfortunately, due to the high luminosity and large pile-up environment, the jet-gap-jet measurement is hard to be reproduced at the LHC. Instead, the so-called jet-veto measurement was proposed \cite{Aad:2011jz}, it amounts to vetoing on the presence of additional jets in the inter-jet rapidity range, instead of strictly imposing a rapidity gap. However, doing so significantly reduces the sensitivity to the BFKL dynamics \cite{DuranDelgado:2011tp,Hatta:2013qj}. In this letter, we propose to take advantage of the future ATLAS Forward Proton (AFP) detector \cite{loi}, in order to realize a true jet-gap-jet measurement. In this context, one would look for jet-gap-jet final states in double-Pomeron-exchange (DPE) processes, with both protons intact. In such events, rapidity gaps are easier to identify. We define a gap to be a rapidity interval devoid of tracks with transverse momentum bigger than 200 MeV, for this reason we shall restrict ourself to central gaps, within the ATLAS tracker acceptance. The modeling of such events can be obtained by implementing the parton-level NLL-BFKL cross section into the Forward Physics Monte Carlo (FPMC) program \cite{Boonekamp:2011ky}.

Compared to the original measurement, using jet-gap-jet final states in DPE processes implies cleaner events, not polluted by proton remnants, since those remain intact. This, in turn, allows to study jets with a larger rapidity difference, for which BFKL effects are supposed to be more important. On the theory side, the fraction of jet-gap-jet to inclusive di-jets events in DPE processes is larger than the corresponding fraction in non-diffractive processes. Indeed, in the former case the gap survival probability factor applies to both the jet-gap-jet and inclusive di-jets events, while in the latter case, it applies only to the jet-gap-jet events.

The plan of the letter is as follows. In section II, we recall the phenomenological NLL-BFKL formulation of the parton-level hard cross section, and we explain how it is embedded into the FPMC program. In section III, we detail the experimental environment for the measurement, provided by the ATLAS experiment and its future forward proton detectors. Our predictions for the DPE jet-gap-jet cross section at the LHC are presented in section IV. Section V is devoted to conclusions and outlook.

\section{Double-Pomeron-exchange jet-gap-jet events in FPMC}

\begin{figure}[t]
\centering
\includegraphics[width=.45\textwidth]{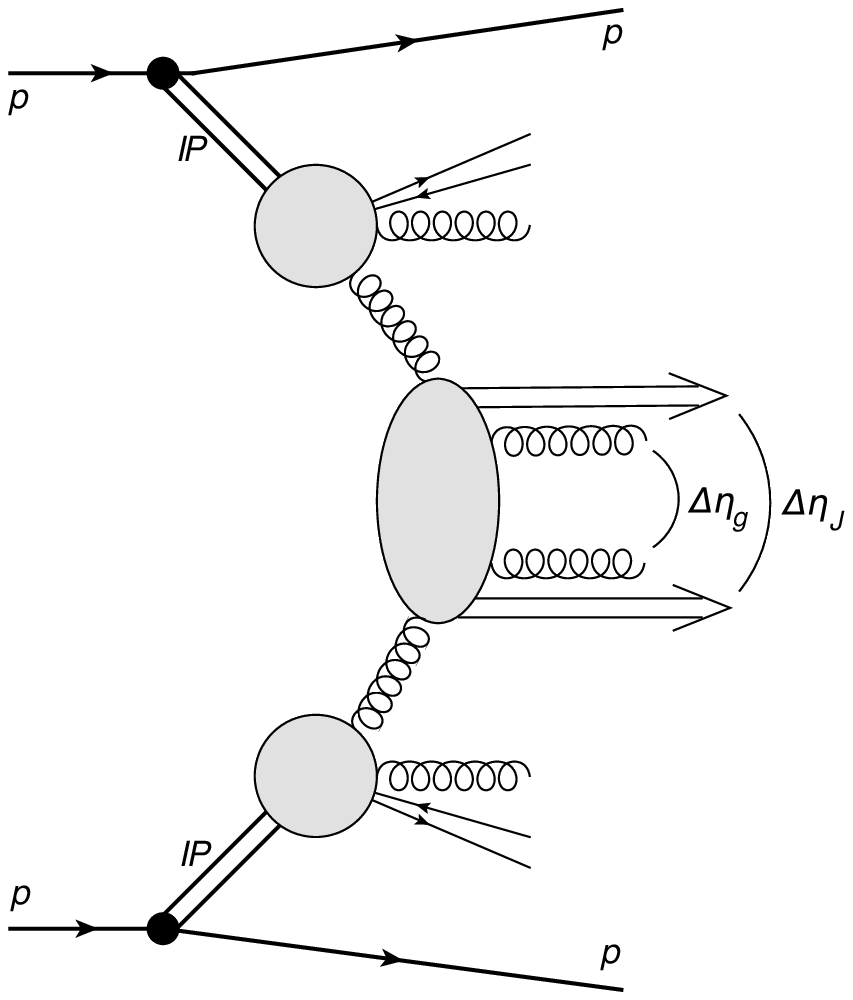}
\hfill
\includegraphics[width=.45\textwidth]{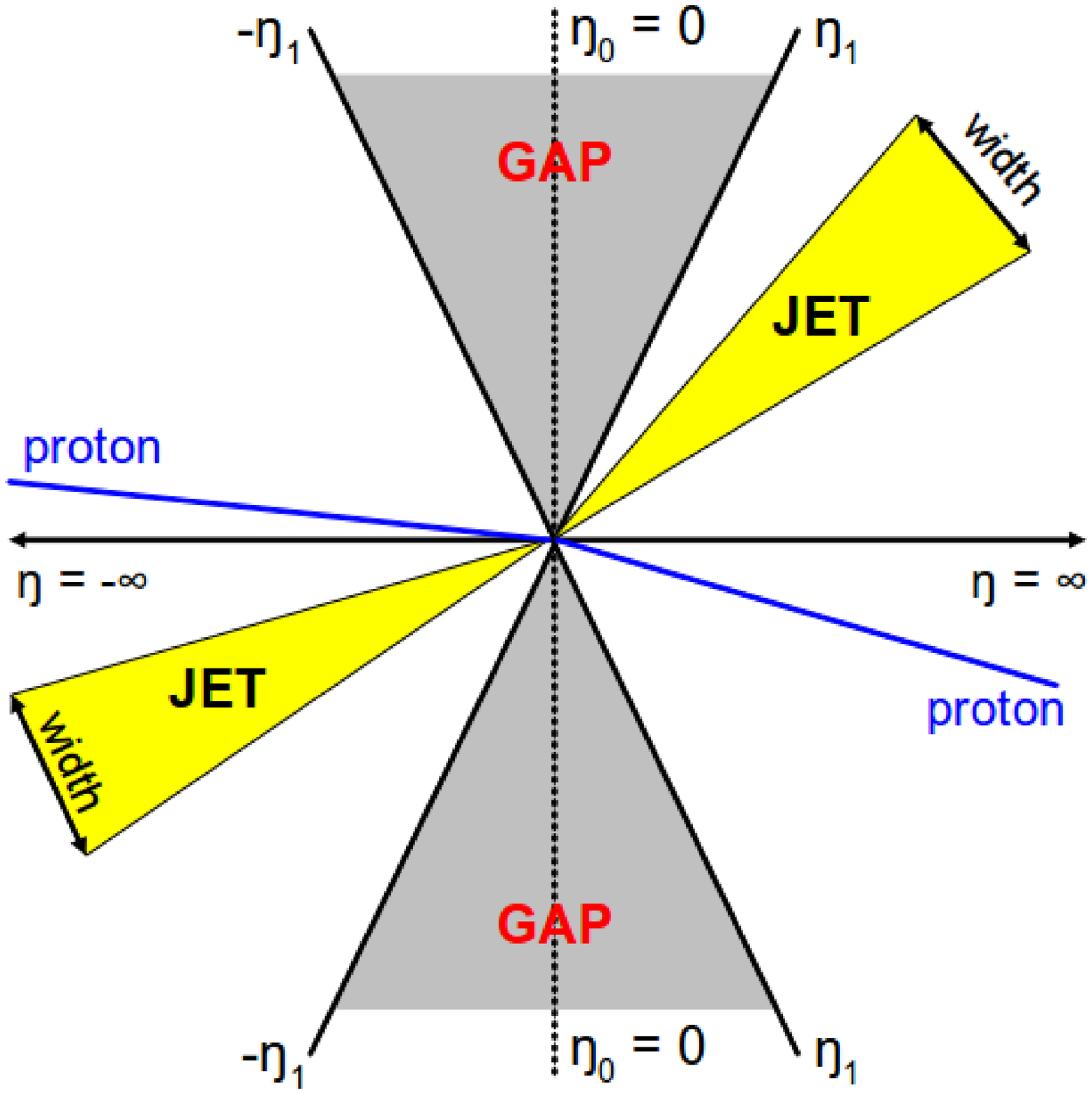}
\caption{
Left: diagram of a DPE event, with a diffractive final state made of two high-$p_T$ jets surrounding a large rapidity gap. $\Delta\eta_{J}$ is the rapidity interval between the jets whereas $\Delta\eta_{g}$ denotes the rapidity gap. Right: the scheme of a DPE jet-gap-jet event in pseudo-rapidity plane.}
\label{fig_feynman}
\end{figure}

To simulate DPE events, the FPMC program is used. This event generator is designed to simulate central particle production with two intact protons, and a hard scale present in the event. In this paper, we deal with the particular case when the centrally produced diffractive final state is made of two high-$p_T$ jets ($p_T\!\gg\!\Lambda_{QCD}$) surrounding a large rapidity gap, as pictured in Fig.~\ref{fig_feynman}.

Practically, in order to implement this specific final state in the FPMC, the function responsible for the calculation of the elementary parton-parton scatterings was modified. The 2-to-2 leading-order matrix elements ordinarily used for the case of central di-jet production, are replaced by 2-to-2 parton-level processes in the BFKL framework. This modification accounts for the $t-$channel exchange of a color-singlet gluon ladder with large momentum transfer, and therefore for the presence of a rapidity gap in between the final-state jets. Of course the collision energy $\sqrt{s}$ should be big ($\sqrt{s}\gg p_T$) in order for jets to be produced along with a large rapidity gap. As detailed below, the leading logarithm (LL) or next-to-leading logarithm (NLL) BFKL kernels are implemented, including all conformal spins and necessary collinear improvements.

\subsection{2-to-2 parton-level processes in the BFKL framework}

In the gluon-gluon channel, the parton-level 2-to-2 hard cross section is given by
\begin{equation}
\frac{d \sigma^{gg\rightarrow gg}}{dp_T^2}=\frac{1}{16\pi}\left|A(\Delta\eta,p_T^2)\right|^2\ ,
\label{hardpart}
\end{equation}
where $A(\Delta\eta,p_T^2)$ is the $gg\to gg$ scattering amplitude. The incoming gluons, of longitudinal momentum fractions $x_1$ and $x_2$ respectively, carry zero transverse momentum, and therefore the final state partons have opposite transverse momenta: $\mathbf{p}_T$ and $-\mathbf{p}_T$. In addition, the rapidity gap coincides with the rapidity interval $\Delta\eta\!=\!\ln(x_1x_2s/p_T^2)$ between the outgoing partons that will initiate the jets. The hadronization of the partons into jets produces a momentum imbalance and reduces the size of the rapidity gap, as is discussed in the next subsection.

Let us now consider the high-energy limit in which the rapidity gap $\Delta\eta$ is assumed to be very large. The BFKL framework allows to compute the $gg\to gg$ amplitude in this regime \cite{Chevallier:2009cu,Kepka:2010hu}:
\begin{equation}
A(\Delta\eta,p_T^2)=\frac{16N_c\pi\alpha_S^2(p_T^2)}{C_Fp_T^2}\sum_{p=-\infty}^\infty\intc{\g}
\frac{[p^2-(\g-1/2)^2]\exp\left\{\bar\alpha(p_T^2)\chi_{eff}[2p,\g,\bar\alpha(p_T^2)] \Delta \eta\right\}}
{[(\g-1/2)^2-(p-1/2)^2][(\g-1/2)^2-(p+1/2)^2]} 
\label{jgjnll}
\end{equation}
with the complex integral running along the imaginary axis from $1/2\!-\!i\infty$ 
to $1/2\!+\!i\infty,$ and with only even conformal spins contributing to the sum.
The running coupling is given by
\be
\bar\alpha(p_T^2)=\f{\alpha_S(p_T^2)N_c}{\pi}=
\left[b\log\lr{p_T^2/\Lambda_{QCD}^2}\right]^{-1}\ ,\quad b=\f{11N_c-2N_f}{12N_c}\ .
\ee
It is important to stress that in order to obtain formula \eqref{jgjnll}, leading-order (non-forward) impact factors were used. In the case that one incoming particle is a quark instead of a gluon, the corresponding amplitude is obtained by multiplying the right-hand side of formula \eqref{jgjnll} by $C_F/N_c$. Note also that the Mueller-Tang prescription \cite{muellertang,leszek} was used to couple the $t$-channel BFKL Pomeron to the incoming colored partons. More details on these parton-level computations, such as the importance of the different conformal spins, or the size of the uncertainties due to the choice of the renormalization scale, can be found in \cite{Chevallier:2009cu}.

In \eqref{jgjnll}, the effective kernel $\chi_{eff}(p,\g,\bar\alpha)$ takes into account NLL-BFKL effects. For $p=0,$ the scheme-dependent NLL-BFKL kernels provided by the regularisation procedure $\chi_{NLL}\lr{\g,\omega}$ depend on $\omega,$ the Mellin variable conjugate to $\exp(\Delta\eta).$ In each case, the NLL kernels obey a consistency condition \cite{salam} which allows to reformulate the problem in terms of $\chi_{eff}(\g,\bar\alpha)$. The effective kernel $\chi_{eff}(\g,\bar\alpha)$ is obtained from the NLL kernel $\chi_{NLL}\lr{\g,\omega}$ by solving the implicit equation $\chi_{eff}=\chi_{NLL}\lr{\g,\bar\alpha\ \chi_{eff}}$. In \cite{nllmnjus,nllmnjthem}, the regularisation procedure has been extended to non-zero conformal spins and the kernel $\chi_{NLL}\lr{p,\g,\omega}$ was obtained from the results of \cite{kotlip}. The formulae needed to compute it can be found in the appendix of \cite{nllmnjus}.\footnote{In the present study we shall use the S4 scheme (scheme 4 of Ref.~\cite{salam}) in which $\chi_{NLL}$ is supplemented by an explicit $\bar\alpha$ dependence. The results in the case of the S3 scheme are similar.} Then the effective kernels $\chi_{eff}(p,\g,\bar\alpha)$ are obtained from the NLL kernel by solving the implicit equation:
\be
\chi_{eff}=\chi_{NLL}\lr{p,\g,\bar\alpha\ \chi_{eff}}\ .
\label{eff}
\ee

We recall that, in the LL-BFKL case, the formula for the jet-gap-jet cross section is formally the same as the NLL one, with the following substitutions in \eqref{jgjnll}:
\be
\chi_{eff}(p,\g,\bar\alpha)\rightarrow\chi_{LL}(p,\g)
=2\psi(1)-\psi\lr{1-\g+\f{|p|}2}-\psi\lr{\g+\f{|p|}2}\ ,
\hspace{1cm}\bar\alpha(p_T^2)\rightarrow\bar\alpha=\mbox{const. parameter} ,
\label{chill}\ee
where $\psi(\g)\!=\!d\log\Gamma(\g)/d\g$ is the logarithmic derivative of the Gamma function. In this case, the coupling $\bar\alpha$ is a priori a parameter, but the preferred value in order to fit the forward-jet data from HERA is 0.16 \cite{nllfjus}. This unphysically small value of the coupling is indicative of the slower Bjorken-$x$ dependence of the forward-jet data compared to the LL-BFKL cross section, when used with a reasonable $\bar\alpha$ value. And in fact, the value $\bar\alpha=0.16$ mimics the slower energy dependence of NLL-BFKL cross section (in this case the average value of $\bar\alpha$ is about 0.25), which in the forward-jet case is consistent with data.

\subsection{Implementation of the NLL-BFKL formula in FPMC}

The parton-level calculation presented in the previous subsection leads by definition to a gap size equal to the interval in rapidity between the partons that initiate the jets. At the particle level, it is no longer true. Due to QCD radiation and hadronization, the jets have a finite size, and the gap size
$\Delta\eta_g$ is smaller than the difference in rapidity between the two jets $\Delta\eta_J$ (see Fig.~\ref{fig_feynman}). The implementation of the NLL-BFKL cross section \eqref{hardpart} into the FPMC Monte Carlo takes these effects into account. Practically, we modified the function which implements the matrix element squared for elementary parton-parton scattering\footnote{This function is named HWHSNM in FPMC, as it was in HERWIG}. Formula \eqref{hardpart}, which gives the BFKL $d\sigma/dp_T^2$ cross section is too complicated to be implemented directly in FPMC since it involves an integration in the complex plane over $\gamma$, and it would take too much computing time to generate a reasonable number of events. To avoid this issue, we parameterized $d\sigma/dp_T^2$ as a function of the parton $p_T$ and $\Delta\eta$ between both partons at generator level. Denoting $z(p_T^2)=\bar\alpha(p_T^2)\Delta\eta/2$, the parametrization used is
\be
\frac{d \sigma}{dp_T^2}=\frac{\alpha_S^4(p_T^2)}{4\pi p_T^4}  \left[ a + b p_T + c \sqrt{p_T}
+ (d + e p_T + f \sqrt{p_T})\times z + (g + h p_T)\times z^2 +
(i + j \sqrt{p_T})\times z^3 + \exp(k + l z) \right]\ .
\label{formulafit}
\ee
This formula is purely phenomenological, not motivated by theory, and was just introduced to obtain a very good $\chi^2$ while fitting \eqref{formulafit} to the full expression of $d\sigma/dp_T^2$. The fit was performed with 12 free parameters $a-l$, and 2330 points were used. Formula \eqref{formulafit} was implemented into FPMC with the fitted parameters.

\section{Experimental Environment}

\begin{figure}[t]
\centering
\includegraphics[width=.95\textwidth]{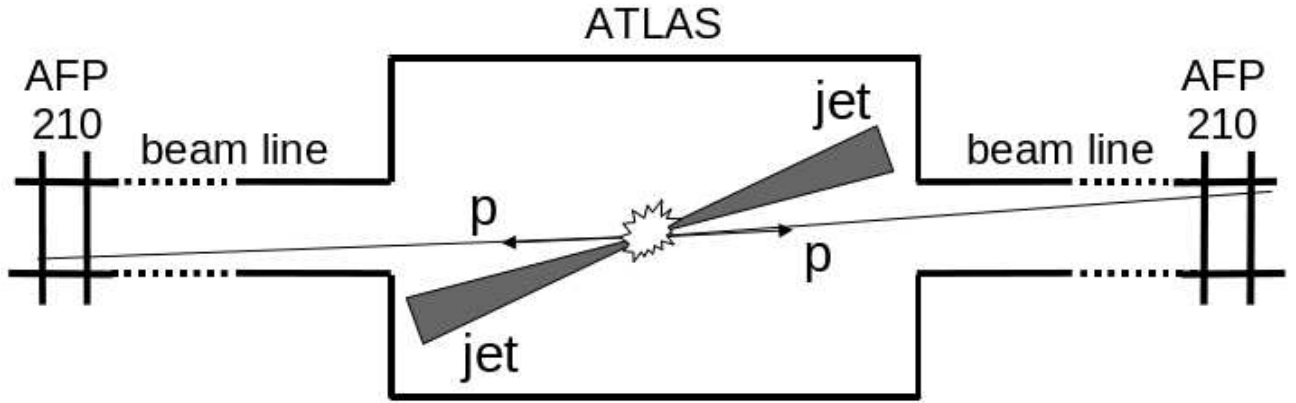}
\caption{Scheme of the measurement concept: jets are registered in the central detectors, and protons in the very forward ones.}
\label{fig_scheme}
\end{figure}

The ATLAS detector is located at the LHC Interaction Point 1 (IP1). It has been designed as a general purpose detector with a large acceptance in pseudorapidity, full azimuthal angle coverage, a good charged particle momentum resolution and a good electromagnetic calorimetry completed by a full-coverage hadronic calorimetry \cite{Aad:2008zzm}. The ATLAS tracking detector provides a measurement of charged particles momenta in the $|\eta| < 2.5$ region and the calorimeter covers $|\eta|<4.9$.

A crucial element of the DPE jet-gap-jet measurement is the possibility to tag the forward protons with the AFP detectors, see Fig.~\ref{fig_scheme}. The AFP detectors are being designed for proton-proton central diffractive or exclusive production measurements \cite{loi}. These detectors will be placed at 204~m and 212~m away from the IP1 inside the ''Hamburg beam-pipes''. These are special devices that allow to put detectors close to the beam and to control the distance between their edge and the proton beam.

\subsection{Forward Protons}

\begin{figure}[t]
\centering
\includegraphics[width=.45\textwidth]{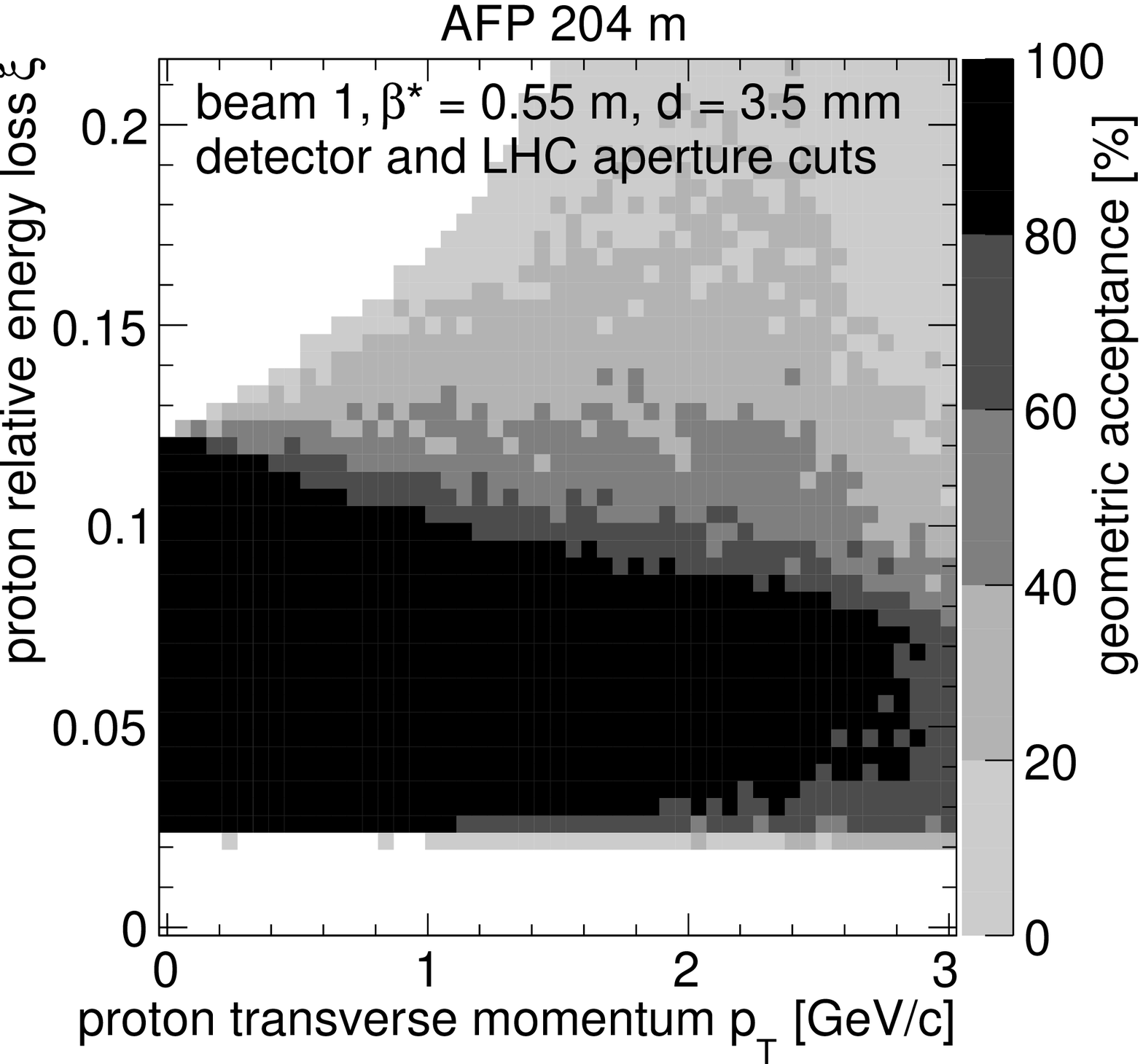}
\hfill
\includegraphics[width=.45\textwidth]{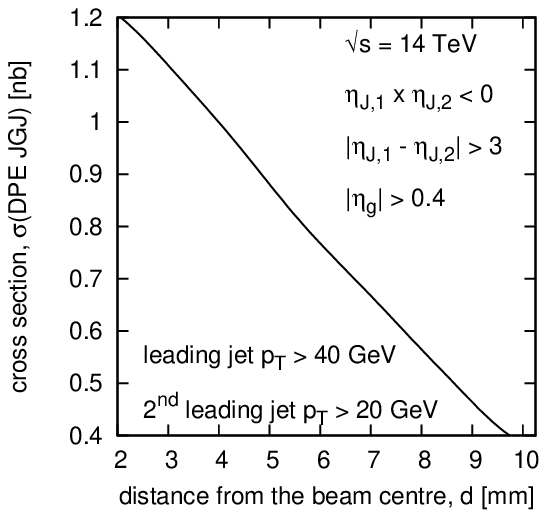}
\caption{Left: geometrical acceptance of the AFP detector as a function of the proton relative energy loss, $\xi$, and its transverse momentum ($p_{T}$). Right: visible cross-section as a function of the distance between the detector and the beam centre (for leading jet with $p_{T} > 40$ GeV). A distance of  3.5 mm would be at 15 sigma from the beam, this is the (standard) value that we choose.}
\label{fig_acceptance}
\end{figure}

Since there are several LHC magnets between the IP1 and the AFP detectors the proton trajectory depends not only on the scattering angle but also on the proton energy. Obviously, not all forward protons can be measured in the AFP detectors. Such protons can be either too close to the beam to be detected or can hit one of the LHC elements (a collimator, the beam pipe) before it reaches the AFP detector. This is included in Fig.~\ref{fig_acceptance}, where the geometric acceptance of the AFP detector is shown. In this calculation the following factors were taken into account: the beam properties at the IP, the beam pipe geometry and the distance between the detector edge and the beam center. As can be observed, the region of acceptance is approximately limited by $0.012 < \xi < 0.14$ and $p_{T} <$ 4~GeV, where $\xi = (1 - E/E_{\mathrm{beam}}$) is the relative energy loss and $p_{T}$ is the proton transverse momentum.

Since both protons need to be tagged in the AFP stations, not all events can be recorded. In fact, the visible cross section depends on the distance between the AFP active detector edge and the beam centre (it will be varied during the runs according to the beam conditions). This dependence is presented in Fig.~\ref{fig_acceptance}~(right). For the rest of the analysis a distance of 3.5~mm is assumed, which results in a visible cross section of about 1~nb (for a leading jet $p_{T} > 40$ GeV).

\subsection{Central Diffractive Jets}

The jets produced in the process under consideration will be measured in the central detector. In the following analysis, the \textit{cone} algorithm with R = 0.7 was used for their reconstruction. To fulfill the ATLAS detector trigger, the leading jet is requested to have a transverse momentum greater than 40 GeV. This rather low value is realistic for the low pile-up LHC runs required (due to the gap reconstruction) to make this measurement. For reconstruction issues, the transverse momentum of the second leading jet is required to be greater than 20 GeV. In this analysis the two leading jets are required to be in the opposite pseudo-rapidity hemispheres and the rapidity gap is required to be symmetric around zero, i.e. from $- |\eta_g|$ to $ |\eta_g|$ with $|\eta_g|=\Delta\eta_g/2$.

\begin{figure}[b]
\centering
\includegraphics[width=.45\textwidth]{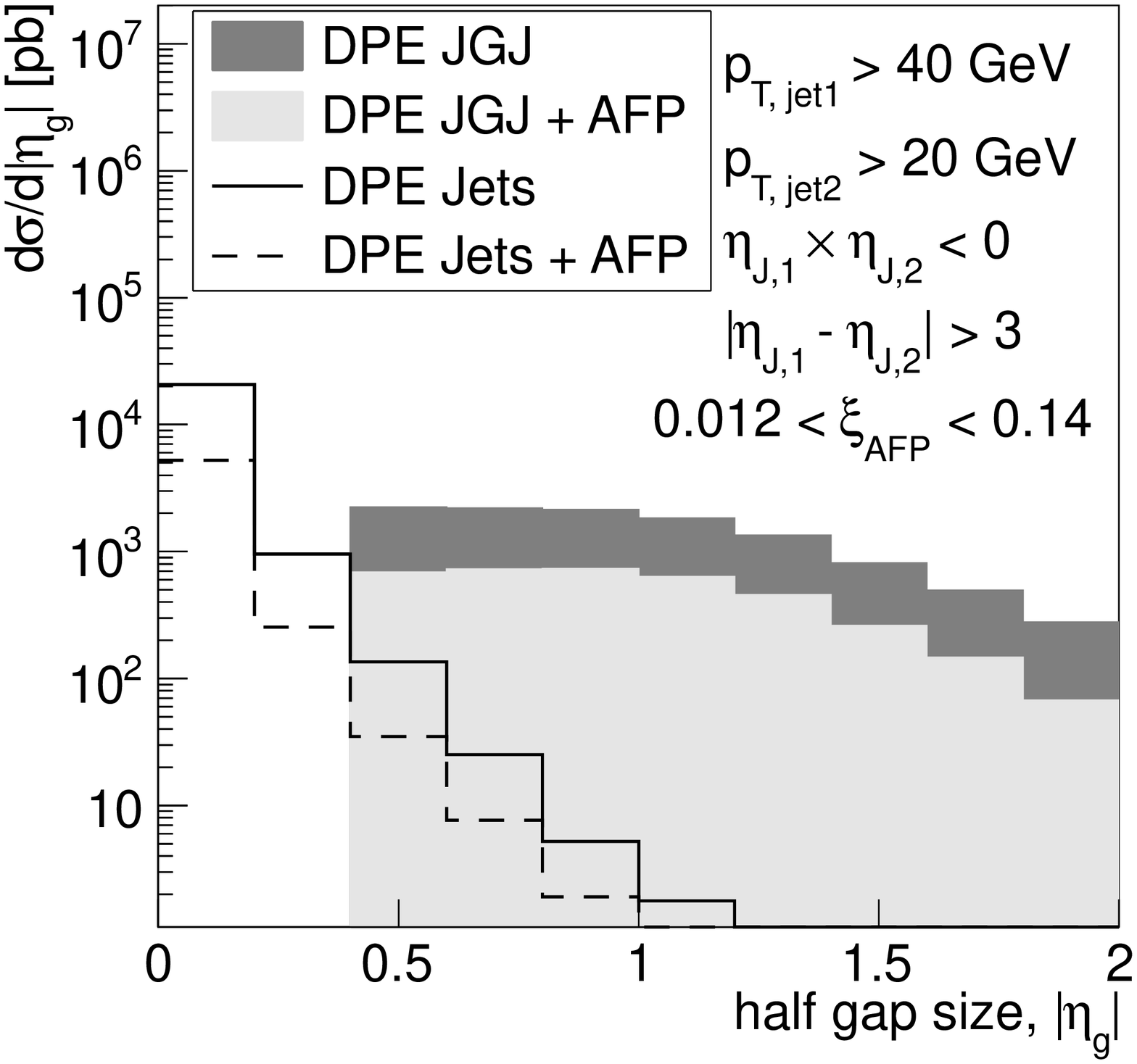}
\hfill
\includegraphics[width=.45\textwidth]{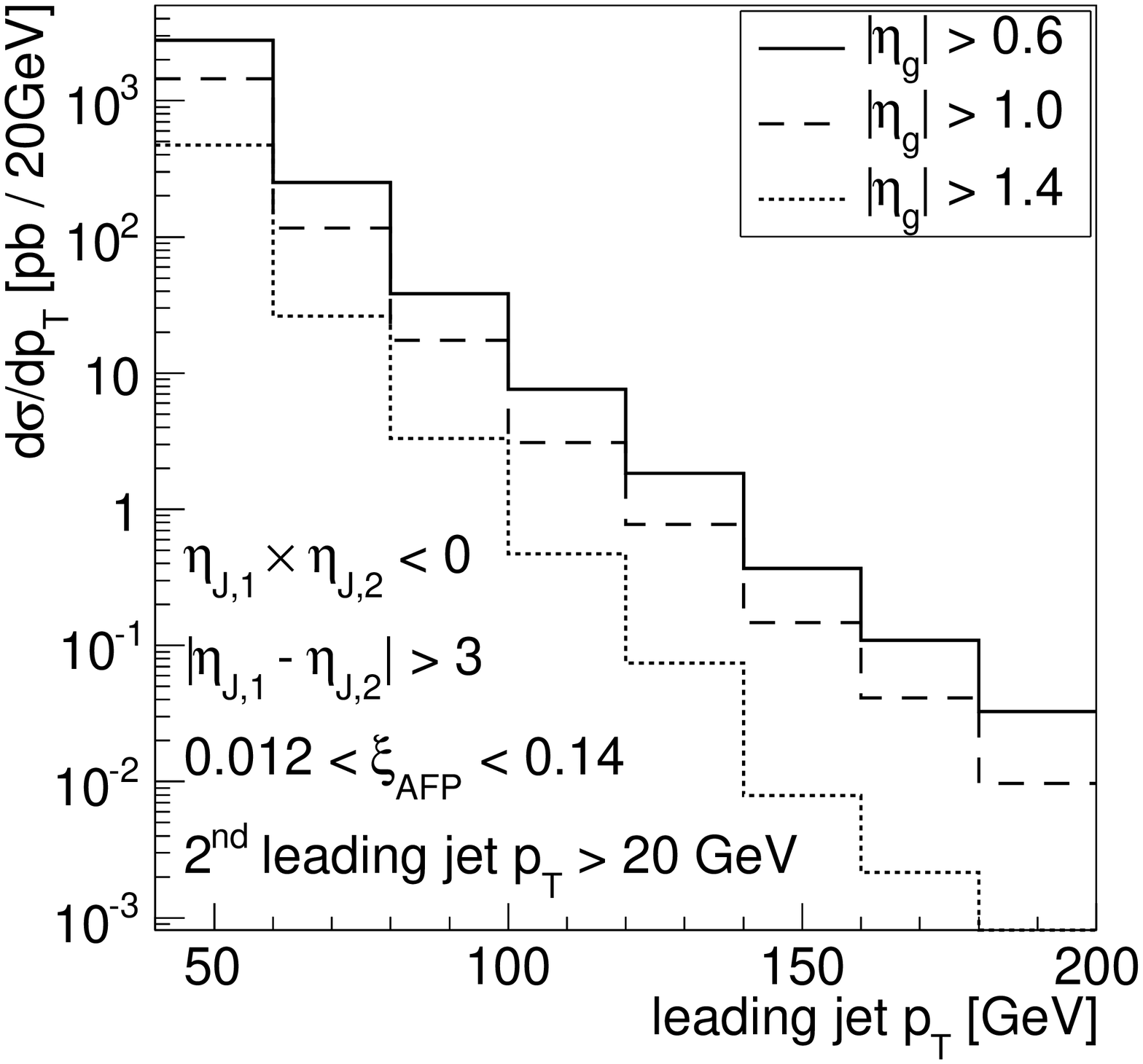}
\caption{Left: the gap size distribution for DPE jets and DPE jet-gap-jet events with and without the AFP tag requirement. We notice that for large enough gaps $\Delta\eta_g>1$, the gap-between-jets events are not dominated by fluctuations in di-jets events. Right: The jet transverse momentum distribution for different gap sizes.}
\label{fig_momentum}
\end{figure}

The main background for the DPE jet-gap-jet production will be DPE inclusive jet production. Indeed, in such processes, a gap between the jets can appear from fluctuations, but this background is significantly reduced by requiring large enough gap sizes (\textit{cf.} Fig \ref{fig_momentum} (left)). Obviously, the bigger the gap size, the larger the DPE jet-gap-jet contribution. However, the cross-section falls steeply. The jet transverse momentum distribution for three different gap sizes ($|\eta_g| =$ 0.6, 1.0, 1.4) is presented in Fig.~\ref{fig_momentum}~(right). Assuming both protons to be tagged in AFP, the good balance between signal to background ratio and visible cross-section is for the gap size $|\eta_g| > 0.5$.

\section{Test of the BFKL Model at the LHC}

\begin{figure}[t]
\centering
\includegraphics[width=.45\textwidth]{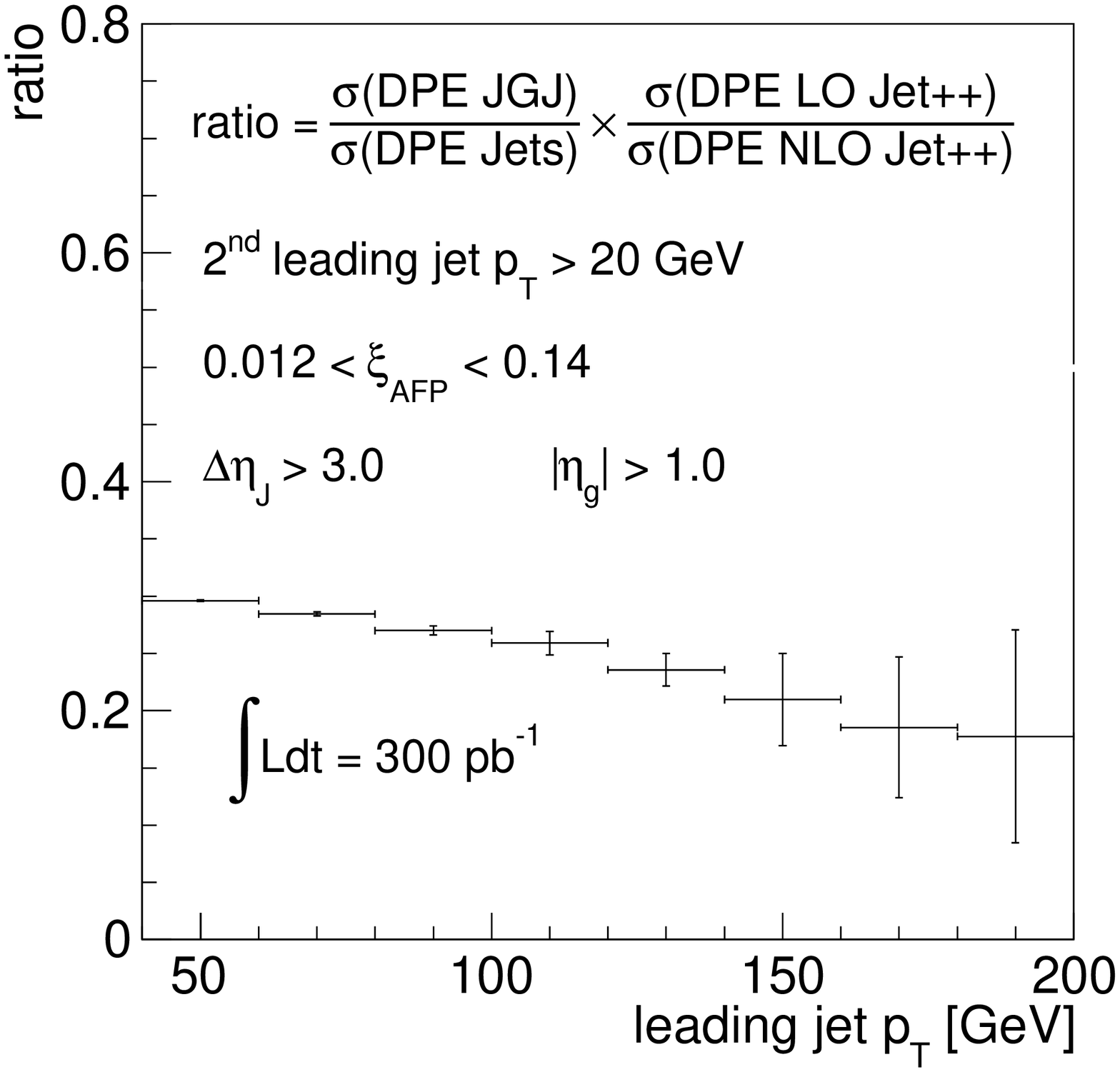}
\hfill
\includegraphics[width=.45\textwidth]{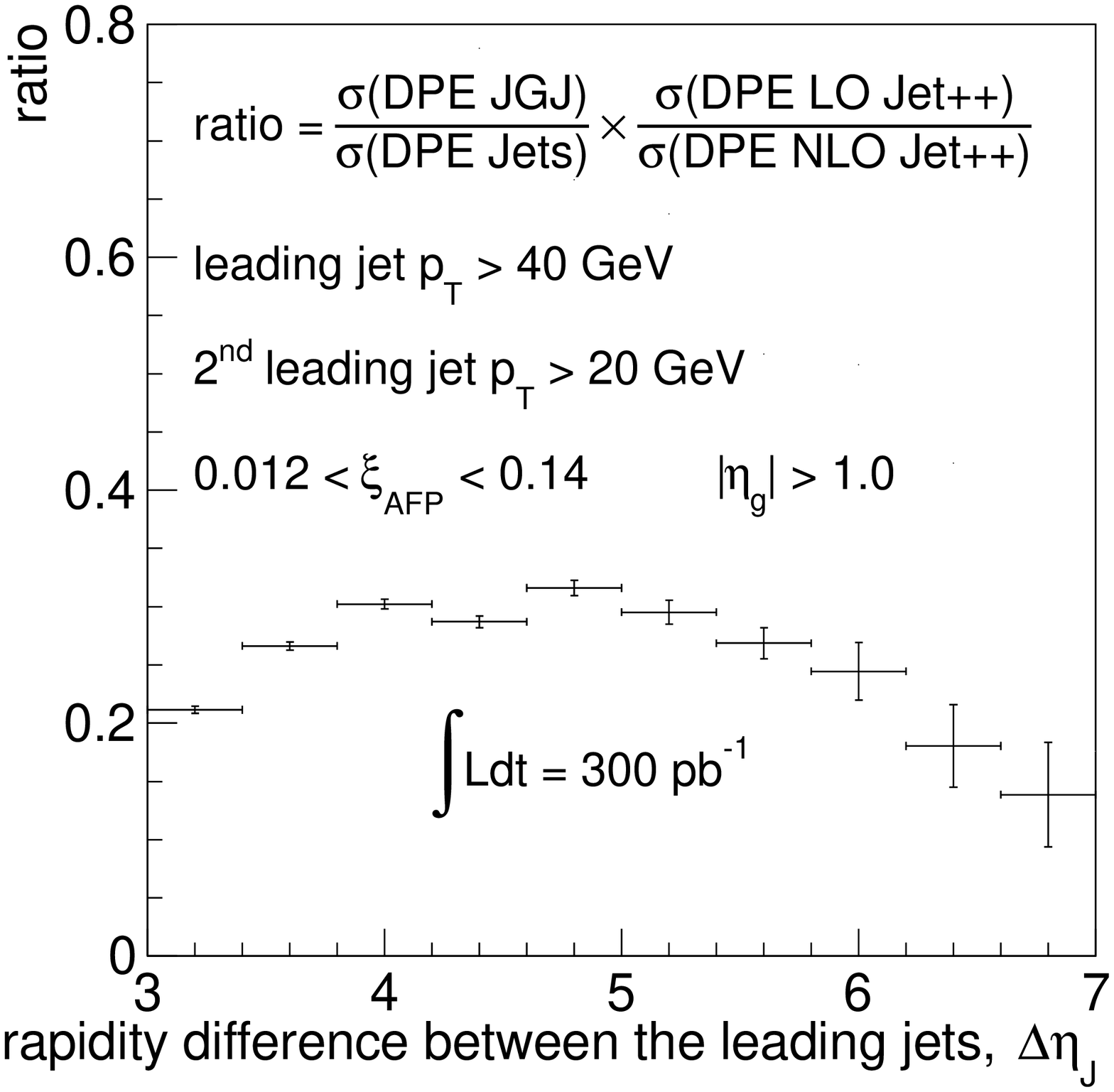}
\caption{Predictions for the DPE jet-gap-jet to DPE jet cross section ratio at the LHC, as a function of the leading jet transverse momentum $p_{T}$ (left), and of the rapidity difference between the two leading jets $\Delta\eta_J$ (right). For both plots, we assumed an integrated luminosity of 300~pb$^{-1}$.}
\label{fig_ratio}
\end{figure}

The DPE jet-gap-jet event ratio is defined as the ratio of the cross section for the DPE jet-gap-jet (JGJ) production to the DPE inclusive di-jet (Jets) production
\be
R = \frac{\sigma ( DPE\ JGJ )}{\sigma ( DPE\ Jets )}\ ,
\label{gap_fraction}
\ee
and is plotted in Fig.~\ref{fig_ratio} as a function of the transverse momenta of the first-leading jet and as a function of the rapidity difference between the two leading jets, $\Delta\eta_{J}$. To verify the power of this test, the statistical errors corresponding to 300~pb$^{-1}$ of integrated luminosity were plotted.

To take into account NLO QCD effects, absent in the FPMC program, the LO ratio obtained was corrected by the cross section ratio $\sigma ( DPE\ LO\ Jet++ )/\sigma ( DPE\ NLO\ Jet++ )$, obtained by the NLO Jet++ program \cite{nlojet}. This program, providing perturbative parton-level cross-sections was adapted for the diffractive applications as described in \cite{Aktas:2007hn} (for di-jet production in p+p collisions, this program was independently validated by comparison with Frixione NLO \cite{Frixione:1997np}). As a crosscheck of the implementation of the resolved Pomeron model into NLO Jet++, the diffractive di-jet cross sections have been successfully compared to the predictions of the Monte Carlo generator RAPGAP \cite{Jung:1993gf}.

For the calculations, diffractive parton distribution functions (DPDF) from the H1 2006 Fit B were used \cite{Aktas:2006hy} (this fit arose from e+p HERA data on inclusive diffraction). The renormalization and factorization scales were set to be equal and identified to the leading jet transverse energy, i.e $\mu_R=\mu_F=E_T^{jet1}$, the jets were reconstructed by the cone algorithm with R=0.7 \cite{Sterman:1977wj}, i.e. by the same method as in FPMC, the number of flavor was fixed to 5, and the strong coupling constant is taken at the 2-loop level with $\alpha_s(M_Z)=0.118$, corresponding to that used in the evolution of the HERA parton densities. Finally, in the NLO  Jet ++ calculations, a rapidity gap survival probability of 0.03 was applied for the LHC. It cancels in the ratio $\sigma ( DPE\ LO\ Jet++ )/\sigma ( DPE\ NLO\ Jet++ )$.

As far as the gap fraction $\sigma ( DPE\ JGJ )/\sigma ( DPE\ Jets )$ is concerned, note that we did not consider an additional suppression factor for DPE jet-gap-jet production, on top of the 0.03 of DPE inclusive jet production. Therefore, in the predictions of Fig.~\ref{fig_ratio}, all the rapidity gap survival probability cancel. We would like to point out that this last point is an assumption, DPE jet-gap-jet production could be subject to a bigger suppression than DPE inclusive jet production, due to the extra color-singlet exchange in the hard cross section. However, we do not expect this potential additional factor (on top of 0.03) to be large, due to the fact that extra soft interactions with the BFKL Pomeron are unlikely: the 2-to-2 hard scattering takes place on much shorter time scale compared to the soft interactions filling the rapidity gaps. In any case, this will be checked at the LHC and if necessary, our numbers can then be adjusted accordingly.

\section{Conclusions}

In order to simulate the production of two high-$p_T$ jets around a large rapidity gap in DPE processes in hadron-hadron collisions, we have embedded the parton-level BFKL calculations of \cite{Chevallier:2009cu} into the FPMC program. The resummations of leading and next-to-leading logarithms are taken into account, and implemented through a renormalization-group improved kernel in the S4 scheme. The Mueller-Tang prescription is used to couple the BFKL Pomeron to colored partons, described with only leading-order impact factors. The implementation of that same NLL-BFKL calculation into the HERWIG Monte Carlo program led to a good description of all Tevatron data on standard jet-gap-jet production \cite{Kepka:2010hu}, which provides a good foundation for this model.

Due to the large pile-up environment, the jet-gap-jet measurement was not reproduced at the LHC, and the model predictions could not be checked. However the measurement can be done in DPE processes, which provides cleaner events not polluted by proton remnants, and consequently also gives access to larger di-jets with a larger rapidity difference, for which BFKL effects are more important. In addition, the fraction of jet-gap-jet to inclusive di-jets events in DPE processes is larger than the corresponding fraction in non-diffractive processes, since in DPE events one is not penalized by the gap survival probability, which applies to both the jet-gap-jet and inclusive di-jet cross section.

After incorporating the parton-level BFKL cross sections into the \textsc{FPMC} generator, we obtained hadron-level results for jet-gap-jet final states in DPE events. In the context of the ATLAS experiment with additional forward physics detectors, we presented our cross section predictions for the LHC, for different experimental settings and gap definitions. They show that there will be enough statistics in 300~pb$^{-1}$ of data to collect DPE jet-gap-jet events. This would provide an additional test of the BFKL Pomeron.

\begin{acknowledgments}

The work of CM is supported by the European Research Council grant HotLHC ERC-2011-StG-279579.
The work of RZ is supported by the grant SVV 265309 of Charles University in Prague.

\end{acknowledgments}

\end{document}